\documentclass[12pt]{article}
\usepackage{geometry}
\usepackage{amsmath}
\usepackage{amsfonts}
\usepackage{natbib}
\usepackage{Sweave}
\usepackage{parskip}

\newcommand{\transpose}{^{\top}}

\renewcommand{\th}{^{\mbox{\tiny th}}}

\title{Computing Robust Leverage Diagnostics when the Design Matrix Contains Coded Categorical Variables}
\author{Kjell Konis}
\date{\today}

\begin{document}

\maketitle

\abstract{For a robust leverage diagnostic in linear regression, \cite{prbz1990} proposed using \emph{robust distance} (Mahalanobis distance computed using robust estimates of location and covariance).  However, a design matrix $X$ that contains coded categorical predictor variables is often sufficiently sparse that robust estimates of location and covariance cannot be computed.  Specifically, matrices formed by taking subsets of the rows of $X$ are likely to be singular, causing algorithms that rely on subsampling to fail.  Following the spirit of \cite{rmvy2000}, we observe that extreme leverage points are extreme in the continuous predictor variables.  We therefore propose a robust leverage diagnostic that combines a robust analysis of the continuous predictor variables and the classical definition of leverage.}

\setlength{\parskip}{\baselineskip}
\setlength{\parindent}{0pt}

\section{Background}

We consider linear regression models of the form
\begin{equation}
y_{i} = x_{i1}\transpose \beta_{1} + x_{i2}\transpose \beta_{3} + x_{i2}\transpose \beta_{3} + e_{i} \qquad (i = 1, \dots, n)
\label{eqn:lm}
\end{equation}
where $x_{i1} \in \mathbb{R}^{p_{1}}$ contains coded categorical predictor variables, $x_{i2} \in \mathbb{R}^{p_{2}}$ contains continuous predictor variables and the elements of $x_{i3} \in \mathbb{R}^{p_{3}}$ are each products of at least one element of $x_{i1}$ and at least one element of $x_{i2}$.  Let $X_{k}$ be the matrix with $i\th$ row $x_{ik}\transpose$ for $k = 1, 2, 3$ so that the design matrix $X = [X_{1} \; X_{2} \; X_{3}]$.  The dimension of $X$ is $n \times p$ where $p = p_{1} + p_{2} + p_{3}$.

Two classical leverage measures are the diagonal elements of the hat matrix (the \emph{hat values})
\begin{equation}
h_{i} = H_{ii} = x_{i}\transpose (X\transpose X)^{-1} x_{i} \qquad (i = 1, \dots, n)
\label{eqn:hat}
\end{equation}
where $x_{i}\transpose = (x_{i1}\transpose \; x_{i2}\transpose \; x_{i3}\transpose)$ is the $i\th$ row of $X$ and the \emph{Mahalanobis distance} (MD)
\begin{equation}
\mbox{MD}_{i} = \sqrt{(x^{*}_{i} - T(X^{*}))\transpose C(X^{*})^{-1} (x^{*}_{i} - T(X^{*}))}
\label{eqn:md}
\end{equation}
where $T(X^{*})$ is the arithmetic mean, $C(X^{*})$ is the sample covariance matrix and $X^{*}$ is identical to $X$ except that the constant column has been removed (if present in $X$).  When $X$ does contain a constant column, these two measures are related by
\begin{equation}
h_{i} = \frac{(MD_{i})^{2}}{n - 1} + \frac{1}{n}.
\label{eqn:himd}
\end{equation}

\section{Robustification}

Let $\{T^{(rob)}, C^{(rob)}\}$ be a robust estimator of location and covariance where the final estimate is a weighted mean and a weighted covariance matrix with weights $w = (w_{1}, \dots, w_{n})\transpose$, $w_{i} \in \{0, 1\}$.  The covariance estimator $C^{(rob)}$ can additionally be rescaled by a factor $c$.  The Fast MCD of \cite{prkd1999} is one such estimator.  The final robust estimate of location is 
$$T^{(rob)}(X_{2}) = \frac{X_{2}\transpose w}{\sum_{i = 1}^{n} w_{i}}$$
and the final robust estimate of covariance is
$$C^{(rob)}(X_{2}) = \frac{c}{(\sum_{i = 1}^{n} w_{i}) - 1} (X_{2} - M)\transpose \, \mbox{diag}(w) \, (X_{2} - M)$$
where $M$ is an $n \times p_{2}$ matrix with rows $[T^{(rob)}(X_{2})]\transpose$.

We then observe that the following modification of $X_{2}$
\begin{equation}
\tilde{X}_{2} = \sqrt{\frac{c (n - 1)}{(\sum_{i = 1}^{n} w_{i}) - 1}} \; W (X_{2} - M) + M.
\label{eqn:mod}
\end{equation}
yields
\begin{equation}
T(\tilde{X}_{2}) = T^{(rob)}(X_{2}) \quad \mbox{and} \quad C(\tilde{X}_{2}) = C^{(rob)}(X_{2}).
\label{eqn:equiv}
\end{equation}

Our idea is to form the \emph{modified design matrix} $\tilde{X} = [X_{1} \; \tilde{X}_{2} \; \tilde{X}_{3}]$ where $\tilde{X}_{3}$ is formed as $X_{3}$ but using the values in $\tilde{X}_{2}$ in place of those in $X_{2}$.  We then define the \emph{robust hat value} to be
\begin{equation}
h_{i}^{(rob)} = x_{i}\transpose (\tilde{X}\transpose \tilde{X})^{-1} x_{i} \qquad (i = 1, \dots, n)
\label{eqn:robhat}
\end{equation}
and the \emph{robust distance} to be
\begin{equation}
\mbox{RD}_{i} = \sqrt{(x^{*}_{i} - T(\tilde{X}^{*}))\transpose C(\tilde{X}^{*})^{-1} (x^{*}_{i} - T(\tilde{X}^{*}))}.
\label{eqn:rd}
\end{equation}

\section{Discussion}

When the linear regression model contains only an intercept term and continuous predictor variables, $X^{*} = X_{2}$, $T(\tilde{X}^{*}) = T^{(rob)}(X_{2})$ and $C(\tilde{X}^{*}) = C^{(rob)}(X_{2})$ so that the quantity defined in equation~\ref{eqn:rd} is equivalent to the robust distance given in \cite{prbz1990}.  Hence, we call this quantity \emph{robust distance} as well.

When $p_{1} > 1$ (i.e., when there are coded categorical predictor variables), the robust distances in equation~\ref{eqn:rd} are appropriate as a leverage diagnostic but not (in the author's opinion) as a distance measure in a multivariate setting.  Therefore we recommend that software report the leverage diagnostic on the scale of the hat values.

\section{Example}

We turn to the epilepsy data published in \cite{ptsv1990} for an example.  

\begin{Schunk}
\begin{Sinput}
> require(robustbase)
> data(epilepsy)
\end{Sinput}
\end{Schunk}

First make the design matrix.

\begin{Schunk}
\begin{Sinput}
> X <- model.matrix(~ Age10 + Base4 * Trt, data = epilepsy)
> n <- nrow(X)
> head(X)
\end{Sinput}
\begin{Soutput}
  (Intercept) Age10 Base4 Trtprogabide Base4:Trtprogabide
1           1   3.1  2.75            0                  0
2           1   3.0  2.75            0                  0
3           1   2.5  1.50            0                  0
4           1   3.6  2.00            0                  0
5           1   2.2 16.50            0                  0
6           1   2.9  6.75            0                  0
\end{Soutput}
\end{Schunk}

In this case we have

\begin{Schunk}
\begin{Sinput}
> X1 <- X[, c(1, 4)]
> head(X1)
\end{Sinput}
\begin{Soutput}
  (Intercept) Trtprogabide
1           1            0
2           1            0
3           1            0
4           1            0
5           1            0
6           1            0
\end{Soutput}
\end{Schunk}

\begin{Schunk}
\begin{Sinput}
> X2 <- X[, 2:3]
> head(X2)
\end{Sinput}
\begin{Soutput}
  Age10 Base4
1   3.1  2.75
2   3.0  2.75
3   2.5  1.50
4   3.6  2.00
5   2.2 16.50
6   2.9  6.75
\end{Soutput}
\end{Schunk}

\begin{Schunk}
\begin{Sinput}
> X3 <- X[, 5, drop = FALSE]
> head(X3)
\end{Sinput}
\begin{Soutput}
  Base4:Trtprogabide
1                  0
2                  0
3                  0
4                  0
5                  0
6                  0
\end{Soutput}
\end{Schunk}

\begin{Schunk}
\begin{Sinput}
> mcd <- covMcd(X2)
> w <- mcd$raw.weights
> mcd$cov
\end{Sinput}
\begin{Soutput}
           Age10      Base4
Age10  0.7463740 -0.3267283
Base4 -0.3267283 10.0194113
\end{Soutput}
\end{Schunk}

The implementation of the Fast MCD in the robustbase package rescales the final covariance matrix estimate by a consistency correction factor \texttt{mcd\$cnp[1]} and a small sample correction factor \texttt{mcd\$cnp[1]} so that $c = \mbox{\texttt{prod(mcd\$cnp)}}$.

\begin{Schunk}
\begin{Sinput}
> cov.wt(X2, wt = w)$cov * prod(mcd$cnp)
\end{Sinput}
\begin{Soutput}
           Age10      Base4
Age10  0.7463740 -0.3267283
Base4 -0.3267283 10.0194113
\end{Soutput}
\end{Schunk}

\begin{Schunk}
\begin{Sinput}
> TX2 <- apply(X2, 2, weighted.mean, w = w)
\end{Sinput}
\end{Schunk}

Compute $\tilde{X}_{2}$ by applying equation~\ref{eqn:mod} to $X_{2}$.

\begin{Schunk}
\begin{Sinput}
> X2.tilde <- sweep(X2, 2, TX2)
> X2.tilde <- sqrt(prod(mcd$cnp)*(n - 1)/(sum(w) - 1) * w) * X2.tilde
> X2.tilde <- sweep(X2.tilde, 2, TX2, FUN = "+")
\end{Sinput}
\end{Schunk}

Verify that $C(\tilde{X}_{2}) = C^{(rob)}(X_{2})$.

\begin{Schunk}
\begin{Sinput}
> var(X2.tilde)
\end{Sinput}
\begin{Soutput}
           Age10      Base4
Age10  0.7463740 -0.3267283
Base4 -0.3267283 10.0194113
\end{Soutput}
\end{Schunk}

We can obtain the modified data (not in general but for this example) by replacing $X_{2}$ in the original data and recomputing the design matrix.

\begin{Schunk}
\begin{Sinput}
> epilepsy[dimnames(X2)[[2]]] <- X2
> X.tilde <- model.matrix(~ Age10 + Base4 * Trt, data = epilepsy)
> head(X.tilde)
\end{Sinput}
\begin{Soutput}
  (Intercept) Age10 Base4 Trtprogabide Base4:Trtprogabide
1           1   3.1  2.75            0                  0
2           1   3.0  2.75            0                  0
3           1   2.5  1.50            0                  0
4           1   3.6  2.00            0                  0
5           1   2.2 16.50            0                  0
6           1   2.9  6.75            0                  0
\end{Soutput}
\end{Schunk}

The final robust leverage measure is then given be the diagonal element of the matrix
$$X (\tilde{X}\transpose \tilde{X})^{-1} X\transpose.$$

\begin{Schunk}
\begin{Sinput}
> diag(X %*% solve(t(X.tilde) %*% X.tilde) %*% t(X))
\end{Sinput}
\begin{Soutput}
         1          2          3          4          5          6          7 
0.05918398 0.05761964 0.07597885 0.08831037 0.12814167 0.03649363 0.05707197 
         8          9         10         11         12         13         14 
0.13821982 0.06977150 0.05953140 0.08231479 0.04790064 0.06109578 0.06518841 
        15         16         17         18         19         20         21 
0.21304208 0.06047114 0.04498471 0.38633944 0.04914452 0.07172279 0.05490496 
        22         23         24         25         26         27         28 
0.09056742 0.05061124 0.04363259 0.06789648 0.12056569 0.10505741 0.07403980 
        29         30         31         32         33         34         35 
0.13316337 0.04489245 0.07575642 0.05223374 0.09433237 0.04382864 0.03457940 
        36         37         38         39         40         41         42 
0.06124138 0.05326251 0.09628077 0.04761239 0.05961493 0.05079567 0.10109938 
        43         44         45         46         47         48         49 
0.06090713 0.05230413 0.06278511 0.06904524 0.03396855 0.05985715 0.64794379 
        50         51         52         53         54         55         56 
0.04181870 0.03780989 0.05743717 0.06796775 0.11009718 0.04673072 0.03927901 
        57         58         59 
0.05935622 0.06818611 0.07601004 
\end{Soutput}
\end{Schunk}

\bibliography{robLev}
\bibliographystyle{plainnat}

\end{document}